\documentclass[12pt,a4paper]{article}
\usepackage{epsfig}
\def\eq#1{{Eq.~(\ref{#1})}}
\def\fig#1{{Fig.~\ref{#1}}}
\def\order#1{$\mathcal{O}{(#1)}$}

\newcommand{\beq}{\begin{equation}}
\newcommand{\eeq}{\end{equation}}
\newcommand{\beqar}[1]{\begin{eqnarray}\label{#1}}
\newcommand{\eeqar}{\end{eqnarray}}
\def\npb#1#2#3{    {\it Nucl. Phys. }{\bf B#1}:#2 (#3)}
\def\plb#1#2#3{    {\it Phys. Lett. }{\bf B#1}:#2 (#3)}

\def\zpc#1#2#3{    {\it Z. Phys. }{\bf C#1}:#2 (#3)}

\def\nc#1#2#3{     {\it Nuovo Cim. }{\bf #1}:#2 (#3)}

\def\sjnp#1#2#3{   {\it Sov. J. Nucl. Phys. }{\bf #1}:#2 (#3)}
\def\jetp#1#2#3{   {\it Sov. Phys. }{JETP }{\bf #1}:#2 (#3)}

\begin{document}
\title {{\bf The Pomeron intercept in $\lambda\phi^3$  theory}\\
{\bf in 4 Minkowski + 1 compact dimensions\\[1cm] }}
\author{
{\bf  
K.~Tuchin\thanks{e-mail: tuchin@post.tau.ac.il} } \\[1cm]
{\it\normalsize HEP Department}\\
{\it\normalsize School of Physics and Astronomy,}\\
{\it\normalsize Raymond and Beverly Sackler Faculty of Exact Science,}\\
{\it\normalsize Tel-Aviv University, Ramat Aviv, 69978, Israel}
}
\date{July, 2000}

\maketitle \thispagestyle{empty}

\begin{abstract}
We calculate the total cross section for two scalar particles scattering
at high energies in $\lambda\phi^3$ theory in five dimensions, four of
which are usual Minkowski ones and the fifth is compact. It is shown that
the cross section is dominated by exchange of Pomeron whose  intercept is
larger than in usual four-dimensional case.
\end{abstract}
\thispagestyle{empty}
\begin{flushright}
\vspace{-14.5cm}
TAUP-2641-2000\\
\today
\end{flushright}

\newpage


\section{Introduction}
The total cross section for various  hadron interactions
at high energy is dominated by the exchange of the Pomeron. 
The Pomeron is defined as a Reggeon (a pole in the scattering
amplitude) with intercept  close to unity. This definition implies  
the following scaling of the total cross section with energy:
\beq\label{11}
\sigma_{tot}=\sigma_0 \left(\frac{s}{s_0}\right) ^\Delta\quad ,
\eeq
where $s$ is the center-of-mass energy squared, $\sigma_0$ and $s_0$ are
certain constants\cite{LE,FR}. $1+\Delta$ is called the Pomeron intercept.
Experimental data can be fitted well with the value
$\Delta=0.08$\cite{DL}. 

In QCD, high energy (leading logarithmic) behaviour of the scattering
amplitude is described by
the solution of the BFLK equation\cite{BFKL}. It  exhibits
Pomeron-like behaviour of \eq{11} with $\Delta=4\ln2\alpha_sN_c/\pi.$
However, there are several problems with this solution. 
First, numerically this  value is much larger than experimentally
observed one. Second, the leading singularity of the scattering amplitude
is a cut rather  than pole, so that actually, the asymptotic behaviour
of $\sigma_\mathrm{tot}$ is more complicated than that of \eq{11}.   
Third, the next-to-leading order corrections
to this solution are so large that they threaten to spoil the perturbative 
approach\cite{NO}. Several authors suggested how to cure  this discrepancy
(see Ref.~\cite{SALAM} for a brief review).
In addition, it was shown in Ref.~\cite{KKL} that the non-perturbative QCD
effects could be responsible for the appearance of the Pomeron pole
(instead of the BFKL cut) in the scattering amplitude.  

In this note we will not be concerned with those problems,
rather we will show that the Pomeron intercept obtains corrections of
completely different origin. 

In recent years the interest arose in theories of Kaluza-Klein type in
which extra compact dimensions are introduced in order to incorporate 
gravity into quantum field theory.  The 4-dimensional
field theory can be thought of as a low-energy effective theory of
graviton excitations and their interactions with matter. 
If the radius of compactification is $R$, then
at energies $s>1/R^2$ certain observables will show dependence on the
scale $R$. It was even argued in the framework of the string theory, that
effects of quantum gravity could be seen at energies on the order of 1~TeV
which will be accessible at LHC\cite{REV}.   
The theories of that kind investigate impact of gravity on the elementary
particles physics. On the
contrary we will assume that the gravity is completely negligible at the
energy scale we are interested in.  If any compact dimensions exist they
will manifest themselves as a corrections to the processes at high
energies even in the flat space-time. We will sum up all such corrections 
for one particular process and show that this leads to increase of the
Pomeron intercept.    
 
In this letter we  discuss a simple model of the Pomeron\cite{MODEL}.
This model is based on a scalar theory with cubic self-interactions. 
Despite its simplicity this model shows many features
of the more realistic QCD Pomeron\cite{LE,FR}. In particular,
the Pomeron of $\lambda\phi^3$ corresponds to the leading  pole in the
scattering amplitude as we 
expect from non-perturbative QCD calculations\cite{KKL}.  

\section{Calculation of the total cross section}
Suppose, for simplicity, that there exist only one compact dimension of
radius $R$. Let us determine what are the corrections to the 
Pomeron of the 4-dimensional $\lambda\phi^3$ theory which stem from this
assumption.
To this end we are going to calculate the high energy
asymptotic of the total cross section for the two scalar particles
scattering.

The $\lambda\phi^3$ theory in  5-dim is defined by the following action:
\beq\label{ACTION5}
S=\int d^4x\int_0^{2\pi R}dy\left(\frac{1}{2}\partial_M\Phi(x,y)
\partial^M\Phi(x,y)-\frac{\lambda_5}{3!}\Phi^3\right)\quad,
\eeq
where $x^\mu$ are the usual  four dimensional space-time coordinates of
Minkowski space and $y$ is a space coordinate of the  compact fifth
dimension. $\Phi(x,y)$ is a self-coupled scalar field, $\lambda_5$ is a
coupling constant. $M=0,1,2,3,4$, $\mu=0,1,2,3$. The metric is flat with
signature $\{1,-1,-1,-1,-1\}$.
Imposing untwisted boundary condition on the scalar field, $\Phi(x,y+2\pi
R)= \Phi(x,y)$ one can  expend $\Phi$ in a Fourier series
\beq\label{FOURIER}
\Phi(x,y)=\frac{1}{\sqrt{2\pi R}}\sum_{n=-\infty}^{\infty}
\phi_n(x)e^{in\frac{y}{R}}=\frac{1}{\sqrt{2\pi R}}
\left[\phi_0(x)+2\sum_{n=1}^{\infty}\phi_n(x)\cos\frac{ny}{R}\right]
\eeq
where we used the fact that $\Phi$ is hermitian operator implying
$\phi_{-n}=\phi_n$.

\begin{figure}\begin{center}
\epsfig{file=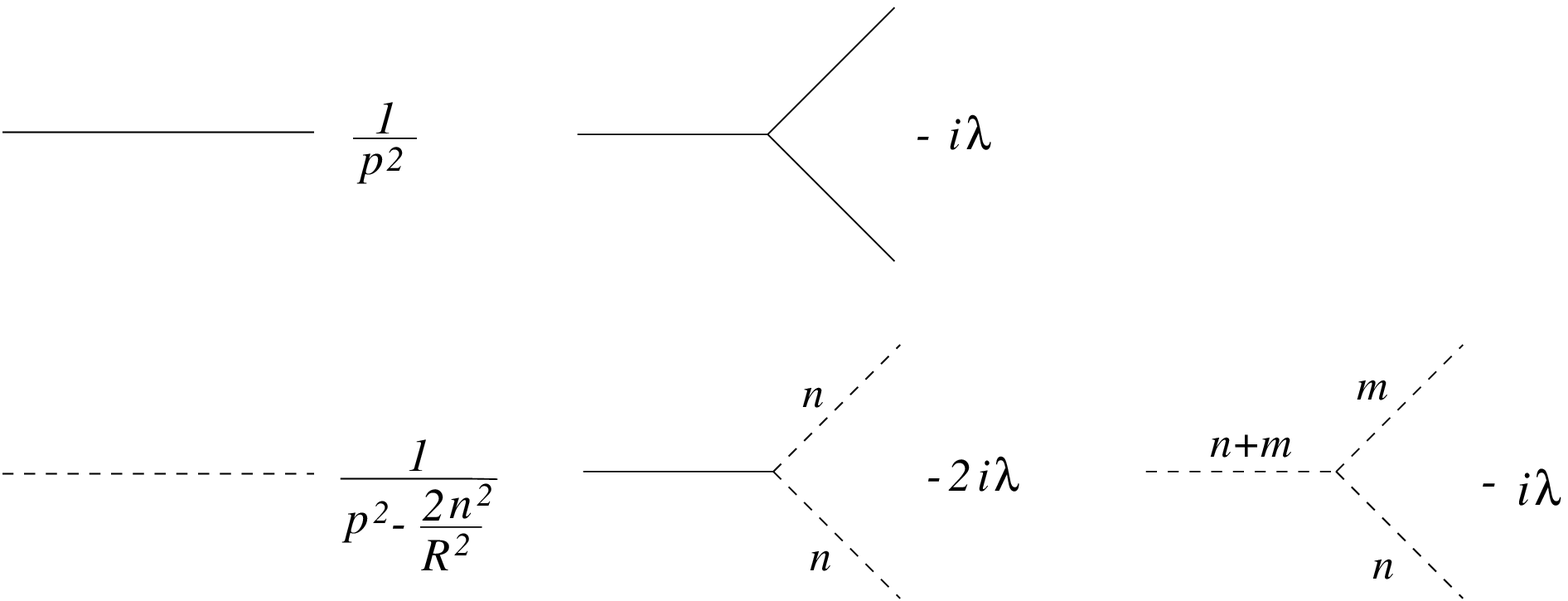, height=5.2cm, width=14.1cm }
\caption{{\it The Feynman rules for the action \eq{ACTION4}. Solid lines
correspond the massless scalar field $\phi_0$, dashed lines -- to massive
scalar fields $\phi_n$. The amplitude must be summed over any free
indecies $n$, $n\ge 1$.}}\label{FEYNMAN}
\end{center}\end{figure}

In order to see how the corrections appear, it is 
convenient to perform calculations in 4-dimensional effective
theory. Upon substitution of \eq{FOURIER} into \eq{ACTION5} and
integration over $y$ we obtain the effective 4-dimensional action which
reads\footnote{The following elementary integrals 
for integer numbers $n,m,p\ge 1$ where used:
\begin{eqnarray*}
\int_0^{2\pi R}dy \cos\frac{ny}{R}&=&0\quad ,\\
\int_0^{2\pi R}dy
\cos\frac{ny}{R}\cos\frac{my}{R}\cos\frac{py}{R}&=&
\frac{\pi}{2}R\left( \delta_{n,m+p}+\delta_{m,n+p}+\delta_{p,m+n}\right)
\quad ,\\
\int_0^{2\pi R}dy \cos\frac{ny}{R}\cos\frac{my}{R}&=&
\pi R\delta_{m,n}\quad ,
\end{eqnarray*}
and the same are for $\sin$.} 
\begin{eqnarray}
S&=&\int d^4 x \left\{ \frac{1}{2}\left(
\partial_\mu\phi_0\right)^2-\frac{\lambda_4}{3!}\phi_0^3\right.\nonumber\\
&&\left.+\sum_{n=1}^\infty\left(\partial_\mu\phi_n\right)^2
-\frac{1}{2}\sum_{n=1}^\infty\frac{2n^2}{R^2}
\phi_n^2-2\frac{\lambda_4}{2!}\phi_0
\sum_{n=1}^\infty\phi_n^2-\frac{\lambda_4}{3!}6\sum_{m,n=1}^\infty
\phi_n\phi_m\phi_{m+n}
\right\},\label{ACTION4}
\end{eqnarray}
where we have defined $\lambda_4=\lambda_5/\sqrt{2\pi R}$. 
The Feynman rules associated with this theory are shown in
\fig{FEYNMAN}. Note, that the field $\phi_0$ corresponds to usual
4-dimensional
theory, while $\phi_n$ are excitations with masses
$m_n^2=2n^2/R^2$. Since $\phi_0$ is massless it the only field
that survives in the low energy limit $s\ll 1/R^2$.

\begin{figure}
\begin{tabular}{ccc}
\epsfig{file=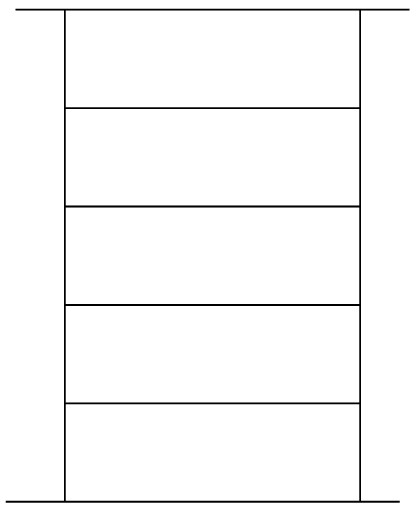,height=3cm,width=2.4cm}&\mbox{~~~~~} &
\epsfig{file=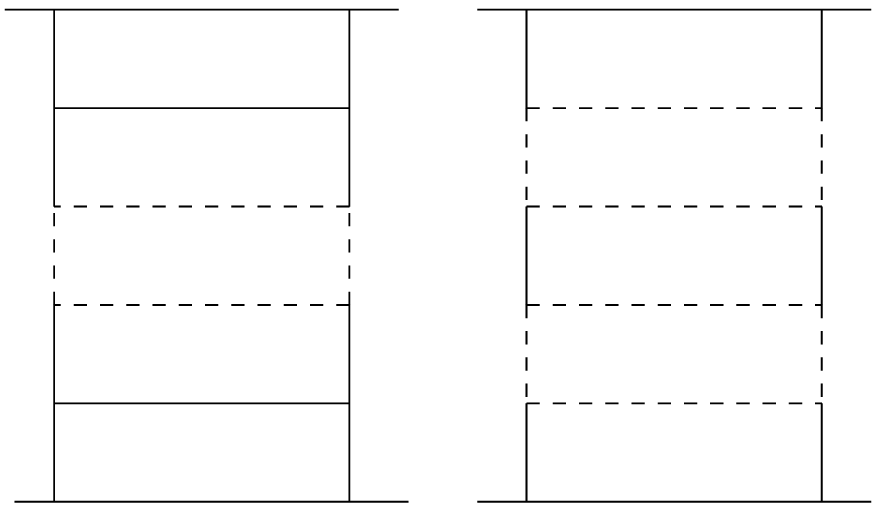,height=3cm,width=5.4cm}\\
(a) & &(b)\\[1cm]
\end{tabular}

\begin{tabular}{ccc}
\epsfig{file=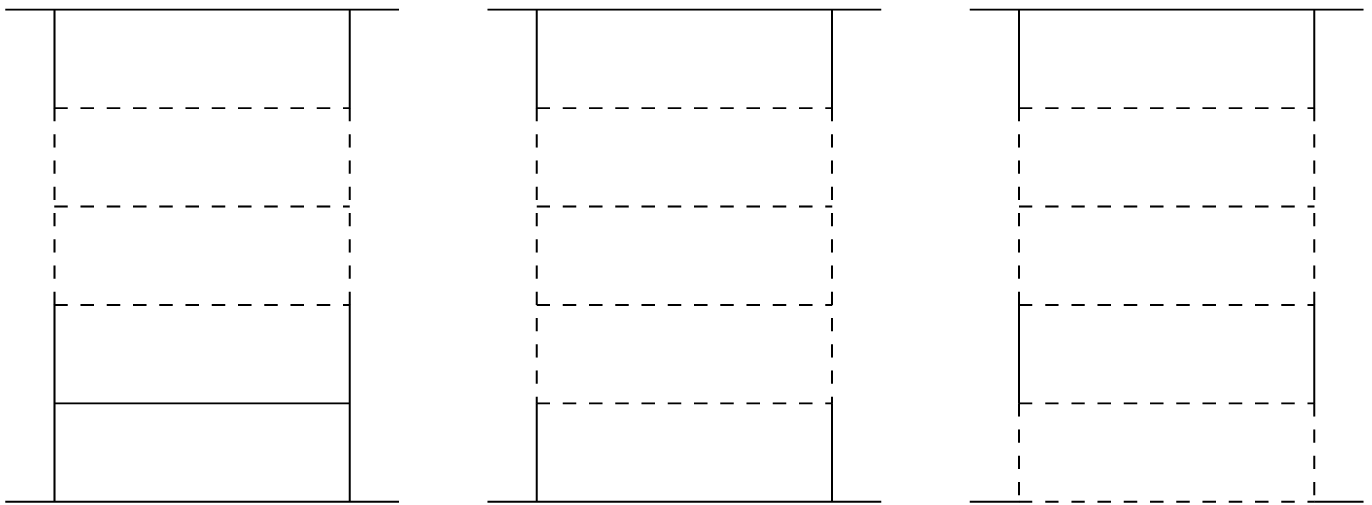,height=3cm,width=8.4cm}&\mbox{~~~~~}&    
\epsfig{file=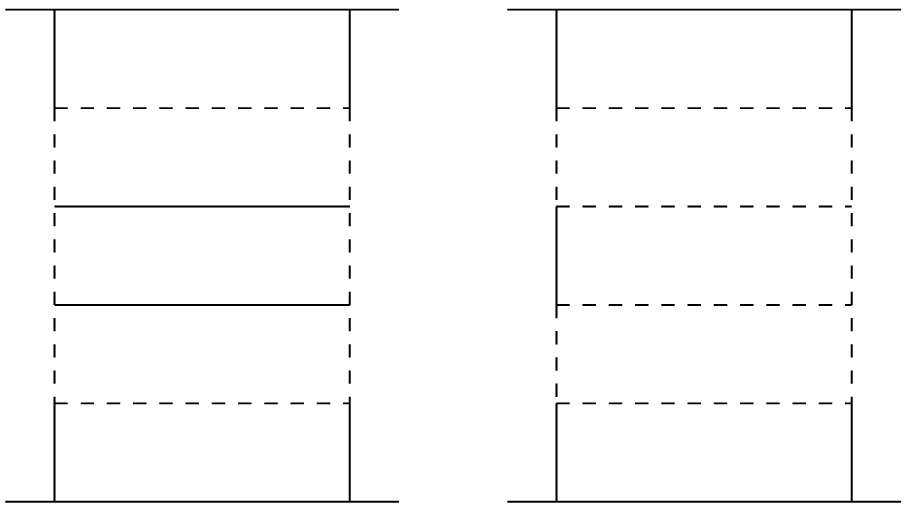,height=3cm,width=5.4cm}\\  
(c) & & (d)
\end{tabular}
\caption{{\it Some ladder diagrams of the order
$\alpha^6$. {\rm (a)}  The ladder of the first kind: no dashed lines
appear, {\rm (b)} ladders of the  second kind: only simple separate
dashed loops appear,  
{\rm (c)} ladders of the third kind: there are merged simple dashed loops
and {\rm (d)} ladders of the fourth kind.
Note, that one can get  ladders of the kind {\rm (d)} by replacing
the dashed lines \emph{inside} merged dashed loops in diagrams of type
{\rm (c)} by the solid ones.    
}}\label{4}
\end{figure}

The dominant contribution to the total cross section for
$\phi_0\phi_0$ scattering in the leading logarithmic approximation
(LLA) ($\alpha\ll 1$, $\alpha\log s\sim 1$) arises from the ladder
diagrams\cite{LE,FR}
In the Fig.~\ref{4} some  diagrams contributing to the order
$\alpha^6$ (four rungs)  are shown ($\alpha=\lambda^2/4\pi$).
We subdivide all ladders at any order of $\alpha$ (fixed number of rungs
$r$) into four kinds. 
\begin{enumerate}
\item 
First, there is the unique ladder without dashes lines (\fig{4}a).
This is the only diagram which contributes in the usual 4-dimensional
$\lambda\phi^3$ theory.
\item 
Second, there are diagrams with \emph{separate simple} dashed
loops, i.e.\  without any two dashed loops having common rung; any
simple loop includes only two neighbor rungs by definition. (\fig{4}b). 
\item  
Third, there are diagrams which                                 
contain at least one pair of \emph{merged} simple dashed loops 
(\fig{4}c). 
\item 
Fourth, all other diagrams (\fig{4}d).
\end{enumerate}

\begin{figure}\begin{center}
\epsfig{file=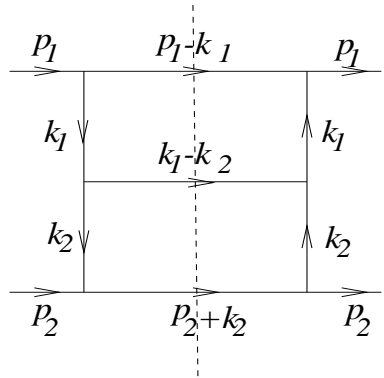,height=5cm, width=5cm}
\caption{{\it Notations for the one rung ladder diagram.
Generalization for arbitrary $r$ is straightforward. 
Dashed line is a $t$-channel cut. 
}}\label{EXA}
\end{center}\end{figure}

Let us calculate the contribution of the $r$-rung diagram of the
first kind. Our notations are defined in \fig{EXA}.
Using optical theorem and Landau-Cutkosky cutting rules  we get
\begin{eqnarray}
\sigma^{r}(\mathrm{1^{st}})&=&
\frac{2}{s}\prod_{i=1}^{r+1}\left(\int\frac{d^4k_i}{(2\pi)^4}\right)
\lambda^2
\prod_{i=1}^{r+1}\left(\frac{\lambda^2}{k_i^4}\right)\nonumber\\
&&\times 2\pi \delta\left((p_1-k_1)^2\right)
\prod_{i=1}^{r}\left( 2\pi\delta\left((k_i-k_{i+1})^2\right)\right)
 2\pi\delta\left((p_2+k_{r+1})^2\right)\label{SIGMA}
\end{eqnarray}
It is convenient to  introduce the Sudakov variables:
$$
k_i = \alpha_i p_1+\beta_i p_2+k_{i\bot}\quad ,
$$
$$
d^4k_i=\frac{s}{2}\: d\alpha_i\: d\beta_i\: d^2k_{i\bot}\quad, i=1,\ldots,
r+1
$$
Leading $\log s$ approximation  stems from the following kinematical
region
$$
\alpha_{r+1}\ll\alpha_r\ll \ldots \ll\alpha_1\ll 1\quad,\quad 
\beta_1\ll\beta_2\ll \ldots \ll\beta_{r+1}\ll 1\quad .
$$
Performing integrations in \eq{SIGMA} in this kinematical region yields
\beq\label{1KIND}
\sigma^{r}(\mathrm{1^{st}})=16\pi^2\alpha \Sigma \frac{1}{s^2}
\Sigma^r\frac{1}{r!}\ln^r\frac{s}{q^2}\quad ,
\eeq
where $q$ is a typical transverse momentum,
\beq
\Sigma=\alpha\int\frac{d^2 k_\bot}{(2\pi)^2}\frac{1}{k_\bot^4}=
\frac{\alpha}{4\pi\mu^2}
\eeq
and $\mu$ is infrared cutoff. Since the integral over $k_\bot$ 
converges at large $k_\bot$ as $1/k_\bot^2$ we set the upper limit of
the integration to be infinity.

The  diagrams of the second kind with given number of rungs
can be drawn by taking the diagram of the first
kind and replacing solid loops by dashed ones in such a way that all dashed
loops are simple and separate. This replacement means, that
\begin{itemize}
\item
every propagator corresponding to the cut dashed line
($t$-channel one) becomes $2\pi\delta(k^2-2n^2/R^2)$.
However, in LLA kinematics contribution of the massless  $t$-channel
propagators is the same as the massive ones. Thus, integration over
the longitudinal Sudakov variables  $\alpha_i$ and $\beta_i$ 
yields the same contribution $\frac{1}{r!}\ln^rs$ as in \eq{1KIND}.
Assume, that the  integration over delta functions is done. Then,
\item
the remaining integration over every dashed loop is performed
over the transverse momentum which replaces the
corresponding solid loop contribution as follows\footnote{Recall
that $k_\bot$ is a space-like 4-vector}:
\beq\label{POPR}
\Sigma=\alpha\int\frac{d^2 k_\bot}{(2\pi)^2}\frac{1}{k_\bot^4}=
\frac{\alpha}{4\pi\mu^2}\longrightarrow
2^4\alpha\sum_{n=1}^\infty\int\frac{d^2 k_\bot}{(2\pi)^2}
\frac{1}{\left(|k_\bot^2|+\frac{2n^2}{R^2}\right)^2}=
\frac{\alpha\,\pi R^2}{3}\quad ,
\eeq
\end{itemize}
The factor $2^4$ arises from the $\phi_0\phi_n^2$ vertex, the coupling of
which is twice as big as $\phi_0^3$ one (see \fig{FEYNMAN}).
This equation implies that every pair of $s$-channel propagators 
gives contribution of the order \order{R^2}  to the cross section.
In general, one can introduce the simple dashed loop into the $r$-rung
diagram of the first kind in $C_{r+1}^{1}=r+1$
ways\footnote{$C_j^i=\frac{j!}{i!(j-i)!}$.}.  
Further, there are $C_{r+1}^k$ ways
to insert $k$ simple dashed loops  into the $r$-rung ladder of the first
kind. It is easy to verify that from those there are 
only\footnote{Note the following relation
$\sum_{l=1}^{k-1}C_{r-l+1}^{k-l}=C_{r+ 1}^{k-1}-1$.}
\beq\label{2WAYS}
 C_{r+1}^k-\sum_{l=1}^{k-1}C_{r-l+1}^{k-l}=C_{r+1}^k-C_{r+1}^{k-1}+1
\eeq
ways to   get the ladder of the second kind ($k\ge 1$, $r\ge 0$).

Next, concentrate on ladders with only simple dashed loops
(\fig{4}b,c) or without them (\fig{4}a). The number of diagrams of
different kinds contributing to the  $r$-rung ladder is given by
\begin{eqnarray}
1 \quad &,& \mathrm{of\, 1^{st}\,\, kind}\nonumber\\
\sum_{k=1}^{\frac{\tilde r}{2}+1}\left(
C_{r+1}^k-C_{r+1}^{k-1}+1 \right) \quad
&,& \mathrm{of\, 2^{nd}\,\, kind}\label{2ND}\\
\sum_{k=\frac{\tilde r}{2}+2}^{r+1}C_{r+1}^k +
 \sum_{k=1}^{\frac{\tilde r}{2}+1}\left( C_{r+1}^{k-1}-1\right)\quad &,&
 \mathrm{of\, 3^{rd}\,\, kind}
          \label{3rd}\label{3RD}\\
\sum_{k=0}^{r+1}C_{r+1}^k=2^{r+1}  \quad &,& \mathrm{altogether}
\label{NUMBERS}
\end{eqnarray}
where 
$
\tilde r=\left\{ {r\qquad,\quad \mathrm{r\,\,  even,} 
\atop r-1\quad,\quad \mathrm{r\,\,  odd.} } \right.
$ 

We will demonstrate now, that the calculation can be significantly 
simplified in the LLA. Consider for example, contribution of the
first and second kind of diagrams. Using \eq{2ND} and \eq{POPR} in
\eq{1KIND} yields
\beq
\sigma(\mathrm{1^{st}+2^{nd}})
 = \frac{4\pi\alpha^2}{s^2\mu^2}
\left\{\sum_{r=0}^\infty
\frac{1}{r!}\ln^{r}\frac{s}{q^2}\,
\left( \frac{\alpha}{4\pi\mu^2}\right)^r
\left[1+ \sum_{k=1}^{\frac{\tilde r}{2}+1}
 \left( \frac{4\mu^2\pi^2 R^2}{3}\right)^k
\left(C_{r+1}^k-C_{r+1}^{k-1}+1
\right) \right]\right\}
 \label{33}
\eeq
Instead of summing over $\sum_{r=0}^\infty\sum_{k=1}^{\frac{\tilde
r}{2}+1}$ it is convenient to
sum in a different order $\sum_{k=1}^\infty\sum_{\tilde r=2k-2}^\infty$\,
 i.e.\ we want  to sum  all contributions to the given
order \order{R^{2k}} first. The key observation is that in the LLA the
following  equation holds
\beq
\sum_{r=j}^\infty\frac{1}{r!}\ln^{r}\frac{s}{q^2}\, a^r r^n\approx
\left(\frac{s}{q^2}\right)^a a^n \ln^n\frac{s}{q^2}\quad ,
\eeq
for any integers $0\le j,n<\infty$. This means that the combinatorial
factor in \eq{33} reads 
\beq\label{NEGL}
C_{r+1}^k-C_{r+1}^{k-1}+1\approx C_r^k\quad .
\eeq  
In particular, the contribution of the ladders of the third and fourth
kinds is
negligible. Indeed, consider \eq{3RD}. Using the symmetry of binomial
coefficients $C_r^k=C_r^{r-k}$ it is easy to rewrite the left hand
term in \eq{3RD} as follows:
\beq\label{UU}
\sum_{k=\frac{\tilde r}{2}+2}^{r+1}C_{r+1}^k=
\sum_{k'=1}^{r-\frac{\tilde r}{2}}C_{r+1}^{r-\frac{\tilde r}{2}-k'}
\quad.
\eeq 
The number of diagrams of the fourth kind (\fig{4}d)
is equal to that given in \eq{3RD} multiplied by a certain function of
the number $k$. From Eqs.~\ref{2ND},\ref{3RD},\ref{NEGL} and \ref{UU} we
see 
that the total contribution of the third and fourth kind ladders
is negligible in comparison with that of the first and second kinds   
at any order \order{R^{2k}}.   
This result justifies our classification of ladder diagrams.

Using \eq{NEGL} in \eq{33} and summing over $r$ and $k$  we obtain
the total cross section in the LLA. It reads
\beq
\sigma=\frac{4\pi\alpha^2}{s^2\mu^2}
\left(\frac{s}{q^2}\right)^{\frac{\alpha}{4\pi\mu^2}
\left(1+\frac{4\pi^2\mu^2 R^2}{3}\right)  }\quad .
\eeq
We thus  see, that the 5-dimensional Pomeron intercept equals
\beq\label{MAIN}
1+\Delta_\mathrm{5dim}=
1+\underbrace{\frac{\alpha}{4\pi\mu^2}}_{\Delta_\mathrm{4dim}}
\left(1+\frac{4\pi^2 \mu^2R^2}{3}\right)
\eeq
i.e. it gets the  positive contribution at energies $s>1/R^2$ as compared
to the 4-dimensional one. 

Our general result, that the fifth space-like compact dimension increases
the value of the Pomeron intercept, is valid in the QCD case as well. 
The positiveness is assured by the integration analogous to that in
\eq{POPR}. Since
the QCD coupling is dimensionless (in 4 dimensions)  correction to the
Pomeron intercept is of the order $\sim+\alpha_S R^2m_\pi^2$. 
So, the QCD Pomeron intercept is equal to 
$1+\Delta_{4\,\mathrm{dim}}$$+c\alpha_S R^2m_\pi^2$,
where $c$ is a certin unknown constant, while $\Delta_{4\mathrm{dim}}$
arises from 4-dimensional theory\cite{BFKL,NO,KKL}. 
Thus, had we known the value $\Delta_{4\,\mathrm{dim}}$, we would have
estimated the size of extra dimension $R$. Meanwile we are used to think
that there are no
manifestations of extra dimensions at accesible energies, i.e.\ 
$R^2 m_\pi^2\ll 1$ in QCD. 
 
To conclude, we studied the influence of the fifth compact dimension on
the Pomeron intercept in $\lambda\phi^3$ theory and found that it gets
positive
correction which is proportional to the squared radius of the compact
dimension. We expect that the same qualitative behavior is valid in QCD. 
This result has impact on both the study of high energy behaviour of the
scattering amplitude  and search of phenomenological signatures of extra
dimensions. 

\vspace{1cm}
{\large\bf Acknowledgments.} 
I wish to thank Professor E.~Levin for numerous  helpful disscussions
and comments as well as for reading the manuscript and providing me with
many refernces.
I would like to acknoledge an interesting discussions with  
S. Bondarenko, Yu. Kovchegov and Ed. Sarkisyan.




\newpage

\end{document}